\documentclass{aa}
\usepackage{graphicx}
\begin{document}
   \title { Misleading results from low-resolution spectroscopy: from
   galaxy interstellar medium chemistry to cosmic star formation density
   {\thanks {Based on observations collected with the ESO Very Large 
   Telescope at the Paranal Observatory (under programmes 66.A-0599(A) 
   and 67.A-0218(A)) and with the Canada-France-Hawaii
   Telescope (under programmes 96IIF25, 98IIF16, 98IF65A and 98IIC14),
   which is operated by CNRS of France, NRC of Canada, and the
   University of Hawaii.}}}

  \titlerunning{Misleading results from low-resolution spectroscopy}

   \author{Y. C. Liang
           \inst{1},
	F. Hammer
	   \inst{1},	 	  
	H. Flores 
	  \inst{1},   
	N. Gruel
	    \inst{2,1},
	F. Ass\'{e}mat
	    \inst{1}             
          }
   \authorrunning{Liang et al.} 

   \offprints{Y. C. Liang,
              email: Yanchun.Liang @obspm.fr;
              F. Hammer,
              email: Francois.Hammer@obspm.fr
          }

\institute{GEPI, Observatoire de Paris-Meudon, 92195 Meudon, France.
      \and
         Department of Astronomy, University of Florida, 
         216 Bryant Space Science Center, P.O. Box
         112055, Gainesville, FL 32611-2055, USA.
      }	      
  
   \date{Received; accepted}

\abstract{Low resolution spectroscopy (R=150) from the
Canada-France-Redshift Survey (CFRS) had revealed intriguing
properties for low redshift galaxies (z$\le$ 0.3): nearly half of
their spectra show prominent $H\alpha$ emission line, but no $H\beta$
emission line and barely detected \ion {[O}{ii]} $\lambda$3727, and
\ion {[O}{iii]} $\lambda$5007 lines. We call these objects ``CFRS
H$\alpha$-single" galaxies and have re-observed a subsample of them at
higher spectral resolution, associated with a subsample of more normal
emission line galaxies.  Good S/N spectroscopy at the VLT and the
CFHT, with moderate spectral resolution (R$>$600), have allowed us to
perform a full diagnostic of their interstellar medium chemistry.
``CFRS H$\alpha$-single" galaxies and most of the star forming spirals
are with high extinctions ($A_V>2$), high stellar masses and
over-solar oxygen abundances.\\ 
From the present study, we believe 
hazardous to derive the detailed properties of galaxies (gas chemical
abundances, interstellar extinction, stellar population, star
formation rates and history) using spectra with resolutions below 600.
One major drawback is indeed the estimated extinction which requires a
proper analysis of the underlying Balmer absorption lines.  We
find that, with low resolution spectroscopy, star formation rates
(SFRs) can be either underestimated or overestimated by factors
reaching 10 (average 3.1), even if one accounts for {\it ad hoc}
extinction corrections.  These effects are prominent for a large
fraction of evolved massive galaxies especially those experiencing
successive bursts (A and F stars dominating their absorption
spectra). Further estimates of the cosmic star formation density at
all redshifts mandatorily requires moderate resolution spectroscopy to
avoid severe and uncontrolled biases.

} 

\maketitle

%

%
\section{Introduction}

Star formation history is a fundamental quantity to study the
populations and evolution of galaxies.  Hot, massive, short-lived OB
stars emit ultraviolet (UV) photons which ionize the surrounding gas
to form an H\,II region, where the recombinations produce spectral
emission lines. Among the Balmer lines, H$\alpha$ is the most directly
proportional to the ionizing UV stellar spectra at $\lambda <$912\AA
$~$(Osterbrock 1989), and the weaker Balmer lines are much more
affected by stellar absorption and reddening.  The other commonly
observed metallic optical lines such as \ion {[N}{ii]}
$\lambda\lambda$6548, 6583, \ion {[S}{ii]} $\lambda\lambda$6716, 6731,
\ion {[O}{ii]} $\lambda$3727, and \ion {[O} {iii]}
$\lambda\lambda$4958, 5007 depend strongly on the metal fraction
present in the gas. Their ionizing potential is higher than the Balmer
lines and thus depends on the hardness of the ionizing stellar
spectra.  These metal lines only represent the indirect tracers of
recent star formation, and they characterize the gas chemistry which
is linked to the star formation history in individual galaxies.
Hence, H$\alpha$ luminosity density is one strong tool to estimate the
cosmic star formation density (CSFD).

The optical H$\alpha$ line has been used to estimate the star
formation (SF) density in the nearby universe.  Tresse \& Maddox
(1998, hereafter TM98) have calculated the H$\alpha$ luminosity
density at z$\sim$ 0.2 on the basis of $z<0.3$ CFRS galaxies and have
obtained 10$^{39.44\pm 0.04}$ ergs s$^{-1}$ Mpc$^{-1}$. More
recent works (Pascual et al. 2001; Fujita et al. 2003), based on deep
imaging narrow band data, provided respectively 1.6 and 1.9 times
higher than TM98 estimated value.  Fujita et al. (2003) argued that
their estimates correspond to a redshift (0.24) slightly higher than
that sampled by TM98, and also that their data revealed a steeper
slope of the faint end of the $H\alpha$ luminosity function.

The above works were based on narrow band filter data and spectroscopy
of very low resolving powers (R from 65 to 150) and they made crude
assumptions about the extinction coefficient to be applied on the
H$\alpha$ luminosity. Indeed, at low resolving power, the underlying
Balmer absorption, related to intermediate age stellar populations,
can severely affect the H$\alpha$/H$\beta$ ratio used to estimate the
extinction coefficient.

In this paper, our aim is to address two questions: (1) could we
estimate the global properties (gas chemical abundances, extinction
and SFRs) of individual galaxies on the basis of low resolution
spectroscopy ?  (2) How solid are SF density estimates based on low
resolution spectroscopy or narrow band filters imagery?  To tackle
these issues we have gathered a small, but representative sample of
low-z galaxies and have systematically compared their properties
derived from different spectral resolution observations, the very low
and the moderate ones.  Low resolution spectroscopy (about 40\AA~) was
provided by the CFRS spectra, and moderate resolution spectra have
been obtained by using the European Southern Observatory (ESO) Very
Large Telescope (VLT) and the Canada-France-Hawaii Telescope (CFHT):
VLT/FORS and CFHT/MOS (5\AA$~$or 12\AA$~$for VLT, 12\AA$~$for CFHT)
  
 The paper is organized as follows. In Sect.2, we describe the sample
selection. The observations, data reduction and flux measurements are
described in Sect.3. Extinction properties are discussed in Sect.4,
allowing to present diagnostic diagrams and the estimated gas
abundances in Sect. 5.  The derived SFRs of these galaxies were given
in Sect.\,6, which includes a discussion on the requirements needed
for proper estimates.  For reasons of consistency with the former
analyses, the adopted cosmological constants are $H_0$=50 km s$^{-1}$
Mpc$^{-1}$ and $q_0=0.5$ throughout this paper.

\section{CFRS low redshift galaxies: the sample selection }
CFRS has produced a unique sample of 591 field galaxies with
$I_{AB}<22.5$ in the range 0$<$\,$z$\,$<$1.3 with a median $\langle z
\rangle \sim 0.56$ (Lilly et al. \cite{CFRS1}, Le F\`evre et
al. \cite{CFRS2}, Lilly et al.  \cite{CFRS3}; Hammer et
al. \cite{CFRSIV}; Crampton et al. \cite{CFRS5}), which is a good
sample to study stellar formation history, stellar population and
evolution of galaxies.

\begin{figure} 
   \includegraphics[bb=21 267 591 706,width=8.8cm,clip]{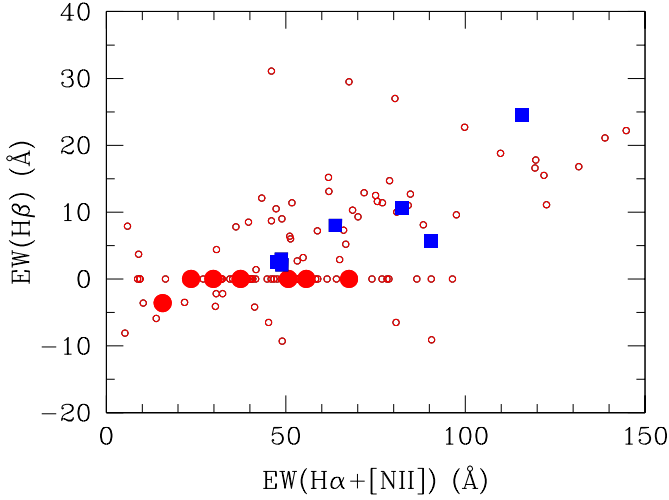}
 \caption{EW(H$\beta$) against EW(H$\alpha$+\ion {[N}{ii]}) for the
117 H$\alpha$ emission CFRS galaxies ($z$$\le$0.3) (open circles)
at rest-frame. Five galaxies are outside the diagram, having higher 
EW(H$\alpha$+\ion {[N}{ii]}) than 150~\AA, 
including two sample galaxies in this 
study, CFRS22.0474 with (370, 42) and CFRS14.1103 with (2940, 192).  
The large filled circles represent the ``CFRS H$\alpha$-single" 
galaxies.  
The filled squares represent the ``CFRS normal emission line" galaxies. }

\label{fig1} 
\end{figure}

There are 138 CFRS galaxies with z $\le$ 0.3 showing both H$\alpha$
and H$\beta$ lines in the rest-frame optical wavelength range. 21 of
these galaxies exhibit both H$\alpha$ and H$\beta$ in absorption, and
117 exhibit H$\alpha$ emission lines (see Fig.~\ref{fig1}).  Among the
117 low-$z$ H$\alpha$ emission galaxies, 57 ($\sim$49\%) of them
exhibit non-positive equivalent width of H$\beta$, EW(H$\beta$)$\leq
0$, (zero in 43 and negative in 14 galaxies), and generally exhibit no
\ion {[O} {iii]} $\lambda 5007$, \ion {[O} {ii]} $\lambda 3727$
emission lines (Hammer et al. \cite{hammer97}; Hammer and Flores
\cite{HF01}; Tresse et al. 1996).  We call these galaxies ``CFRS
H$\alpha$-single" galaxies.  53 of the H$\alpha$-emission line
galaxies show EW(H$\beta)>0$, which are called as ``CFRS normal
emission line" galaxies in this study.  The other 7 are without
available information about H$\beta$ because of the weak quality of
their spectra near to 4861\AA $~$wavelength (Tresse et al. 1996;
Hammer et al. 1997).

Why almost half of the CFRS low-$z$ sample galaxies exhibit
``H$\alpha$-single" spectra ?  Could it be due to large extinctions?
Is it related to the low spectral resolution in the CFRS ?  Or could
it be due to the ``fact" that they are ``peculiar" objects?  What is
the difference or relation between the ``H$\alpha$-single" galaxies
and other ``normal emission line" galaxies?

Seven ``CFRS H$\alpha$-single" galaxies are selected to understand
their detailed properties (the filled circles on
Fig.~\ref{fig1}).  Another nine ``CFRS normal emission line" galaxies
(showing both H$\alpha$ and H$\beta$ in emission) are associated to be
selected for comparison mainly (the squares on Fig.~\ref{fig1}).  They
can be the representatives of the two group galaxies mentioned above
as demonstrated in Fig.~\ref{fig1} though all of them were observed
during observational runs targetting different goals (higher redshift
galaxies) by completing MOS or FORS masks.

The properties of the 16 sample galaxies are studied by using
moderately high resolution and high S/N spectra from the VLT and the
CFHT.  Their images were obtained by the Hubble Space Telescope (HST)
WFPC2 with filter F814W in 1994 and the CFHT/FOCAM in 1991.  The basic
data for these galaxies are given in Table~\ref{tab1}. The columns are
CFRS name, redshift, $I_{AB}$ magnitude, absolute B magnitude,
$K_{AB}$ and absolute K magnitudes (``9999" was marked for the absent
values), and the telescopes used for spectral and imaging observations.
All the magnitudes are in the AB systems.

{
\begin{table} 
{ 
 \tiny
\caption { Basic data of the sample galaxies, the top seven are the ``CFRS H$\alpha$-single"
galaxies, the bottom nine are the ``CFRS normal emission line" galaxies. 
``SPE" means ``spectroscopy" and ``IMA" means ``imagery"}  
\label{tab1}

\begin{tabular}{cccccccccc} \hline

Objects  & $z$   & $I_{AB}$    & $M_B$   & $K_{AB}$ &  $M_K$  &  SPE &  IMA  \\ \hline
03.0364  & 0.2511 & 19.05 &$-$20.79  & 17.89   &  $-$22.63 &      CFHT  &   HST \\ 
03.0365  & 0.2183 & 19.19 &$-$20.09  & 17.91  & $-$22.20 &     CFHT  &   HST  \\ 
03.0578  & 0.2192 & 20.79 &$-$19.16  & 20.07  & $-$20.04 &  VLT600 &    HST  \\
03.0641  & 0.2613 & 20.03 &$-$19.62  & 9999  & 9999  &  VLT600 &    CFHT \\
03.0711  & 0.2615 & 21.04 &$-$19.03  & 19.76  & $-$20.92  &    VLT600 &    HST  \\ 
03.1014  & 0.1961 & 18.42 &$-$20.87  & 17.27  &   $-$23.10 &      CFHT &    HST  \\  
22.0717  & 0.2791 & 19.60 &$-$20.24  & 17.92   &  $-$23.04  &   VLT300 &    CFHT \\ \hline

03.0003  & 0.2187 & 22.49 & $-$16.60 & 21.81    &    $-$18.18 &    CFHT  &   CFHT \\ 
03.0149  & 0.2510 & 20.74 & $-$19.42 & 19.70   &  $-$20.82  &   VLT600 &    HST    \\    
03.0160  & 0.2184 & 21.83 & $-$17.71 & 9999    & 9999  &    CFHT  &   CFHT \\ 
03.1299  & 0.1752 & 18.59 & $-$20.47 & 9999   & 9999  &     CFHT  &   HST \\ 
03.1311  & 0.1755 & 19.56 & $-$19.40 & 18.05   & $-$22.23  &     CFHT  &   HST \\ 
14.1103  & 0.2080 & 22.33 & $-$17.69 & 9999   & 9999  &     CFHT  &   CFHT \\ 
14.1117  & 0.1919 & 20.79 & $-$18.98 & 9999   & 9999  &     CFHT  &   CFHT \\ 
22.0474  & 0.2801 & 21.74 & $-$18.66 & 21.12   & $-$19.55  &   VLT300 &    CFHT     \\
22.1084  & 0.2930 & 20.29 & $-$20.22 & 19.22   & $-$21.70  &     CFHT  &   CFHT \\   \hline 
 \end{tabular}
}
\end{table}
}

\section{Observations, data reduction and flux measurements}

Spectrophotometric observations for four sample galaxies were obtained
during one night (for 3$^h$ field) with the ESO 8m VLT/FORS2
instrument with the R600 and I600 grisms (R=5\AA) and covering the
wavelength range from 5500 to 9200\AA.  The slit width was
1.2{$^{\prime\prime}$}, and the slit length 10$^{\prime\prime}$.  The
objects CFRS22.0474 and CFRS22.0717 were observed in another night
(for 22$^h$ field) using the VLT/FORS2 instruments with the R300 grism
(R=12\AA) and covering the wavelength range from 5800 to 9500\AA.  The
slit width was 1.0\arcsec, and the slit length 10$^{\prime\prime}$.
CFHT data were obtained in different runs from 1996 to 1999, using
the standard MOS setup with the R300 grism and a 1.5$^{\prime\prime}$
slit.  These ensure the coverage from H$\beta$ to \ion {[S}{ii]} lines
in the rest-frame spectrum.

The spectra were extracted and wavelength-calibrated using
IRAF{\footnote{IRAF is distributed by the National Optical
Astronomical Observatories, which are operated by the Association of
Universities for Research in Astronomy, Inc., under cooperative
agreement with the National Science Foundation.}}  packages.  Flux
calibration was done using 15 minute exposures of different
photometric standard stars. To ensure the reliability of the data, all
spectrum extractions, as well as the lines measurements, were
performed by using the SPLOT program.

\begin{figure*} 
   \includegraphics[bb=52 95 538 802,width=16.8cm,clip]{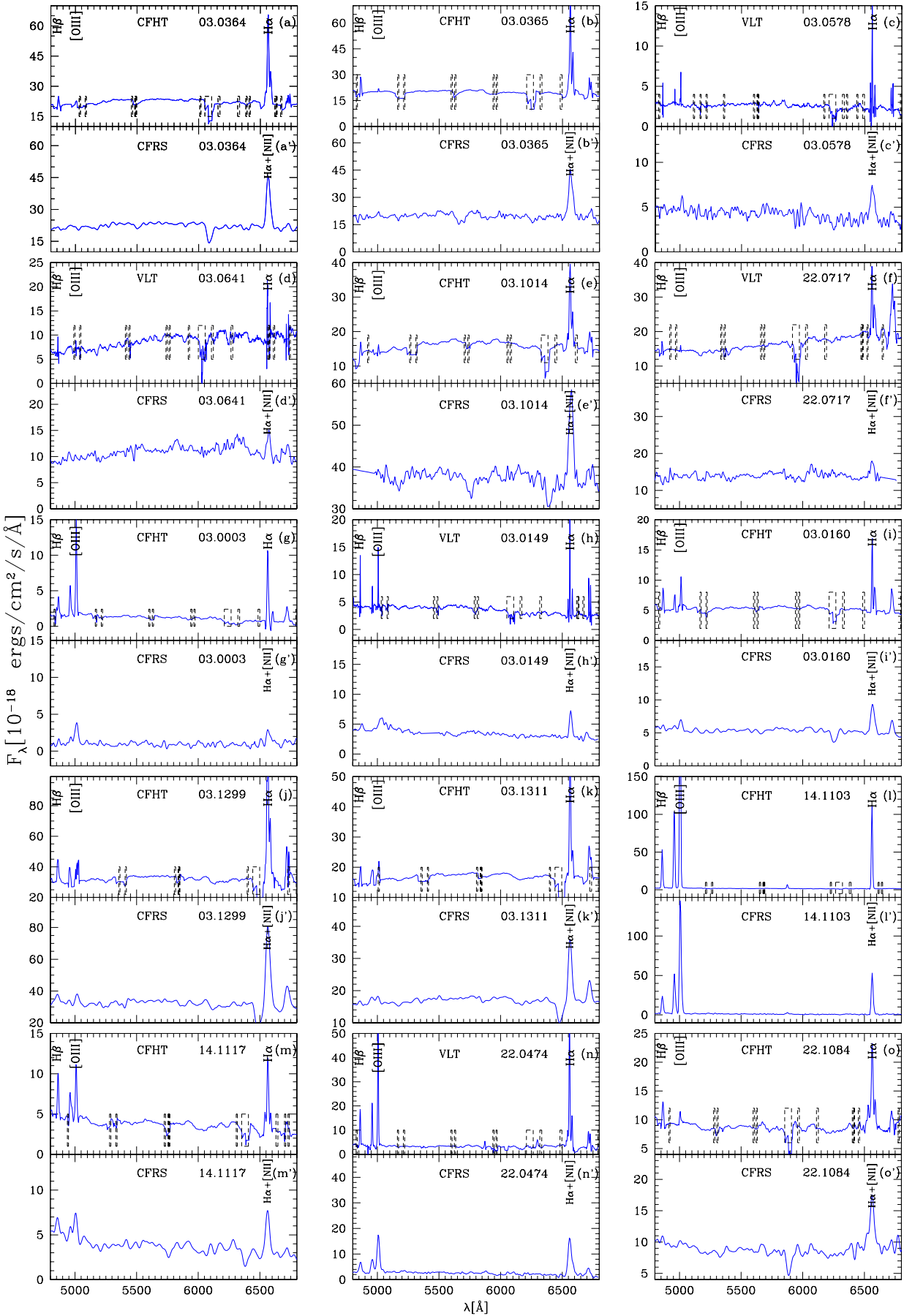}

\caption{{\bf (a)-(o)} Rest-frame spectra of 15 sample galaxies from
the moderate-resolution observations ; {\bf (a$\arcmin$)-(o$\arcmin$)}
the corresponding low-resolution CFRS spectra of the galaxies.  {\bf
(a)-(f)} and {\bf (a$\arcmin$)-(f$\arcmin$)} are the spectra of the
six ``CFRS H$\alpha$-single" galaxies, others are the spectra of the
nine ``CFRS normal emission line" galaxies.  The spectrum of another
``CFRS H$\alpha$-single" galaxy, CFRS03.0711, will be given in
Fig.~\ref{fig3}.}
 \label{fig2}
\end{figure*}

\begin{figure} 
   \includegraphics[bb=22 171 567 692,width=8.8cm,clip]{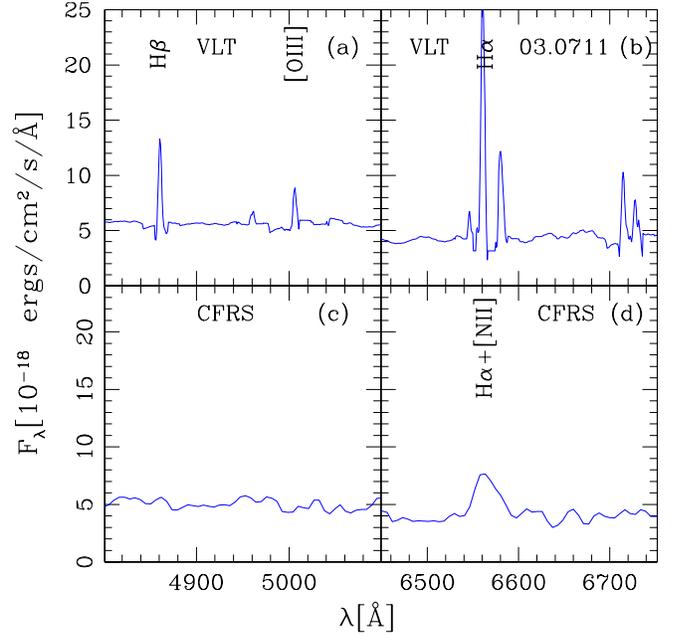}
\caption{{\bf (a), (b)} Rest-frame spectra of CFRS03.0711 from the VLT
observation around H$\beta$ and H$\alpha$ wavelength positions; 
{\bf (c), (d)} The corresponding low-resolution CFRS 
spectra.}
 \label{fig3}
\end{figure}

The rest-frame spectra of the 16 sample galaxies are given in
Fig.~\ref{fig2}{\bf (a-o)} and Fig.~\ref{fig3}{\bf (a),(b)}.  The
continua have been convolved except at the locations of the emission
lines (e.g. H$\beta$; \ion {[O}{iii]} $\lambda\lambda$4958, 5007;
\ion{[N}{ii]} $\lambda\lambda$6548, 6583; H$\alpha$ and \ion{[S}{ii]}
$\lambda\lambda$6716, 6731) using the procedure developed by our
group (Hammer et al. \cite{hammer01}; Gruel \cite{Gruel02}; Gruel et
al. \cite{Gruel03}).  For the VLT600 spectra, the adopted
convolution factors are 7 pixels and then 15 pixels; for the VLT300
and CFHT spectra, the convolution factors are 7 and 7 pixels (Hammer
et al. \cite{hammer01}; Gruel et al. \cite{Gruel03}).  Pairs of
vertical dashed lines delimit the regions where strong sky emission
lines (e.g. \ion{[O}{i]} 5577, 6300, 6364\AA$~$and OH 6834, 6871,
7716, 7751, 7794, 7851, 7914, 7993, 8289, 8298, 8344, 8505\AA) and
absorption lines (O$_2$ 6877, 7606 and 7640\AA) are located.
Fig.~\ref{fig2}{\bf (a$\arcmin$-o$\arcmin$)} and Fig.~\ref{fig3}{\bf
(c),(d)} give the corresponding CFRS low-resolution spectra of the
galaxies.  The comparison between the moderate-resolution spectra and
the CFRS spectra show that the emission lines are strongly hidden or
diluted in the low resolution observations.  And the higher resolution
make it possible to separate the \ion{[N}{ii]} $\lambda\lambda$6548,
6583 emission lines from the H$\alpha$ emission.

The fluxes of emission lines have been measured using the SPLOT
package.  The stellar absorption under the Balmer lines are estimated
from the synthesized stellar spectra obtained using the stellar
spectra of Jacoby et al. (\cite{Jacoby84}).  The corresponding error
budget has been deduced using three independent methods: the first one
is estimated by trying several combinations of the stellar templates
for the stellar absorption; the second one is from measurement, which
is estimated according to the independent measurements performed by
Y.C. Liang, H. Flores and F. Hammer; the third one is Poisson noises
from both sky and objects.
The flux measurements of the emission lines from the VLT and the CFHT
spectra and their errors are given in Table~\ref{tab2}.  The fluxes of
\ion{[O}{ii]}$~\lambda3727$ are estimated from the original CFRS
spectra with large error bars due to the absence of this line in the
rest-frame wavelength ranges of the moderate resolution spectra.

{
\begin{table*} 

{  
\tiny
\caption [] {Measured emission fluxes (F$_\lambda$) of the sample
galaxies in units of 10$^{-17}$ ergs cm$^{-2}$ s$^{-1}$, ``9997" means
the line is blended with strong sky line, ``9998" means there are no
the corresponding lines detected around the line positions, and
``9999" means the lines are shifted outside the observed wavelength
ranges.  ``Ap" is the aperture correction factor between the CFRS and
the present moderate-resolution observations due to the different slit
widths}
 
\label{tab2}


\begin{tabular}  {rrrrrrrrrrrr} \hline
 CFRS  & H$\beta$ 4861 &  \ion{[O}{iii]}4958 & \ion{[O}{iii]}5007 
  & H$\alpha$6563 &\ion{[N}{ii]}6548  &\ion{[N}{ii]}6583  
  & \ion{[S}{ii]}6716 & \ion{[S}{ii]}6731  & \ion{[S}{ii]}$_{1+2}$  & \ion{[O}{ii]}3727 
  &  Ap  \\ 
  
  &  &  &  
  &  &  &   
  & &  & &  
  &    \\  \hline
  
03.0364   & 7.18$\pm$0.57   & 0.81$\pm$0.05    & 2.44$\pm$0.14 
          &  62.30$\pm$0.62   & 6.89$\pm$0.41     & 22.3$\pm$1.11
          &  --- & ---    & 10.90$\pm$1.50   & 9999
          & 1.0          \\    [1.5mm]

03.0365  &13.49$\pm$0.54  & $<$0.54          & $<$1.61 
         & 88.75$\pm$8.34 &14.25$\pm$1.43  & 33.14$\pm$2.45 
         &  12.55$\pm$1.04  &  9.32$\pm$0.77 & 21.87$\pm$1.81 & 14.0$\pm$14.0
         &  1.0       \\ [1.5mm]

03.0578  & 1.23$\pm$0.20  & 0.61$\pm$0.09 & 1.77$\pm$0.26 
         & 8.76$\pm$0.43 & 1.06$\pm$0.34 & 1.80$\pm$0.66 
         & 1.80$\pm$0.31   & 2.07$\pm$0.29 & 3.87$\pm$0.60 & 5.3$\pm$1.0 
         & 1.8    \\   [1.5mm]

03.0641  & 2.21$\pm$0.24 & 0.31$\pm$0.06& 0.92$\pm$0.17
         & 10.25$\pm$0.92  & 1.18$\pm$0.14 & 3.53$\pm$0.43
         &$<$1.38    &$<$1.65   & $<$3.03     & 3.0$\pm$1.0  
         &  1.3   \\ [1.5mm]

03.0711  & 4.49$\pm$0.18 & 0.65$\pm$0.07  & 1.95$\pm$0.20
         & 13.58$\pm$0.58 & 1.04$\pm$0.12 & 5.78$\pm$0.68
         & 3.56$\pm$0.37 & 1.96$\pm$0.37  & 5.52$\pm$0.74 & 2.5$\pm$2.0 
         & 1.0      \\ [1.5mm]

03.1014  & 5.90$\pm$0.71& 9998   & 1.32$\pm$0.26 
         & 43.28$\pm$4.51& 4.24$\pm$0.85  & 13.00$\pm$1.71  
         & 9.26$\pm$0.99 & 5.12$\pm$2.78 & 14.38$\pm$3.77& 16.0$\pm$7.0 
         & 2.2   \\           [1.5mm]

22.0717  & 3.54$\pm$0.44  & 9998    &$<$0.75 
         & 29.71$\pm$2.67& 3.67$\pm$0.40   & 14.07$\pm$1.55
         & --- & --- & 9997 & 1.3$\pm$1.3
         &  1.0      \\ [1.5mm]

  \hline

03.0003  & 3.97$\pm$0.80  & 5.41$\pm$0.54  & 16.16$\pm$0.86 
         &  13.89$\pm$0.62    & ---  & $<$3.06  
         & 2.80$\pm$0.20  & 1.95$\pm$0.14  &  4.75$\pm$0.50  &  2.1$\pm$2.1 
         & 0.6   \\  [1.5mm]

03.0149  & 4.76$\pm$0.28 & 1.87$\pm$0.07 & 5.43$\pm$0.22
         &17.18$\pm$0.71 & 1.35$\pm$0.17  & 3.49$\pm$0.43 
         & 3.95$\pm$0.72  & 2.61$\pm$0.39 & 6.56$\pm$1.11 & 7.1$\pm$7.1 
         & 1.0   \\  [1.5mm]

03.0160    &4.25$\pm$0.51  & 1.57$\pm$0.157    &4.72$\pm$0.47
           &12.40$\pm$0.37  & --- & 6.17$\pm$0.66
           & --- & ---  &7.04$\pm$0.68  &9999 
           &1.0     \\  [1.5mm]

03.1299   &16.4$\pm$1.48   & 5.06$\pm$0.46   &15.2$\pm$1.37
          &   144.0$\pm$4.32 & 13.2$\pm$1.32  & 39.6$\pm$3.96
          & --- & ---    & 40.0$\pm$4.80  & 9999
          & 1.0     \\  [1.5mm]

03.1311  & 6.97$\pm$0.49  &  2.53$\pm$0.20  &  7.59$\pm$0.61 
         & 58.4$\pm$2.34  &  4.40$\pm$0.18   & 13.4$\pm$0.54
         & 16.5$\pm$1.65 & 10.5$\pm$1.05 & 27.0$\pm$2.7  & 9999 
         & 1.0       \\  [1.5mm]

14.1103   & 60.56$\pm$0.61 &  130.9$\pm$0.52   & 392.7$\pm$1.57  
          & 133.9$\pm$0.54    & 9998   & 9998 
          & --- & ---  & 9998   & $<$3.45   
          & 1.0        \\  [1.5mm]

14.1117   & 5.54$\pm$0.44     & 3.34$\pm$0.20  & 10.03$\pm$0.60 
          & 15.9$\pm$0.48  & ---  & 2.70$\pm$0.54  
          & --- & ---   & 4.08$\pm$0.28   & 9999
          & 1.0     \\ [1.5mm]

22.0474  &  13.04$\pm$0.46 &  14.05$\pm$0.30  &  43.70$\pm$1.20  
         & 51.97$\pm$2.49   &   4.20$\pm$0.40  &  11.97$\pm$2.0
         & 6.90$\pm$0.68  & 5.80$\pm$0.58  &  12.70$\pm$1.26  &  17.0$\pm$2.3
         &  1.0    \\  [1.5mm]

22.1084   & 5.23$\pm$0.52 &  0.94$\pm$0.12 & 2.82$\pm$0.37  
          & 26.9$\pm$2.15 & 1.70$\pm$0.23   & 5.11$\pm$0.71  
           & --- & ---  &  5.82$\pm$0.64   & 9999 
          & 1.0        \\  \hline
 \end{tabular} 

}

{\scriptsize Notes: Ap=$({{\rm CFRS}\over{\rm VLT}})_{\rm Aper}$. 
   \ion{[S}{ii]}$_{1+2}$=\ion{[S}{ii]}${6716}$+\ion{[S}{ii]}${6731}$. 
   }
\end{table*} 
}

\section{Balmer decrement and extinction}
The major factor that affects measurements of the true emission fluxes
is interstellar extinction. If galaxies are observed at high galactic
latitudes, the extinction due to our own Galaxy is negligible
($\sim$0.05 mag), hence most extinction is intrinsic to the observed
galaxy.  Extinction arising along the line of sight to a target galaxy
makes the observed ratio of the flux of two emission lines differ from
their ratio as emitted in the galaxy. The extinction coefficient, $c$,
can be derived using the Balmer lines H$\alpha$ and H$\beta$:

\begin{equation}
\frac 
{I(H\alpha)}{I(H\beta)}=\frac{I_0(H\alpha)}{I_0(H\beta)}10^{-c[f(H\alpha)-f(H\beta)]},
\end{equation}

where $I(H\alpha)$ and $I(H\beta)$ are the measured integrated line
fluxes, and $I_0(H\alpha)/I_0(H\beta)$ is the ratio of the fluxes as
emitted in the interstellar dust. Assuming case B recombination, with
a density of 100\,cm$^{-3}$ and a temperature of 10$^4$\,K, the
predicted ratio of $I_0(H\alpha)$ to $I_0(H\beta)$ is 2.86 (Osterbrock
1989).
 
\begin{figure}
   \includegraphics[bb=19 146 436 686,width=8.8cm,clip]{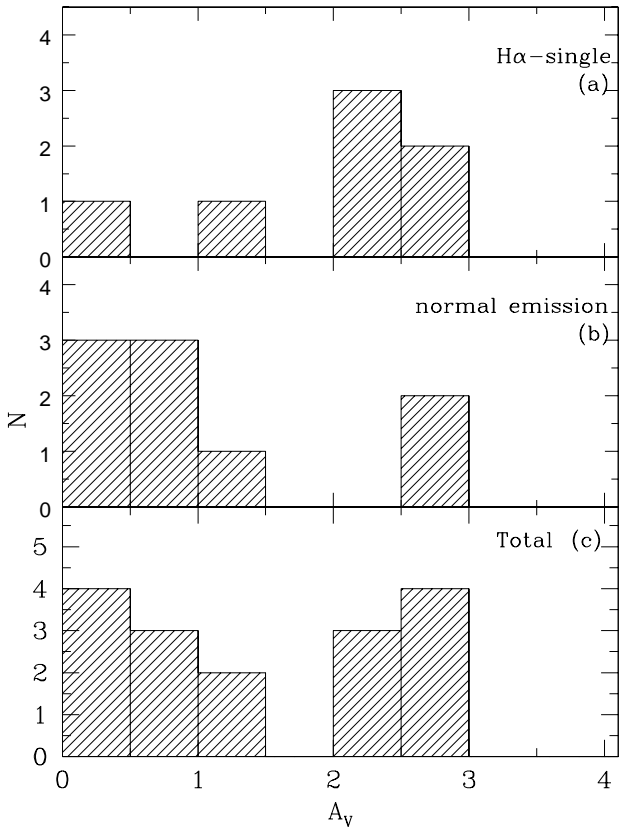}
   \caption{Distribution of the extinction $A_V$ in sample galaxies
   with a binning interval of 0.5: {\bf (a)} the seven ``CFRS
   H$\alpha$-single" galaxies; {\bf (b)} the nine ``CFRS normal
   emission line" galaxies; {\bf (c)} the total 16 sample galaxies. }
\label{fig4} 
\end{figure}

Using the average interstellar extinction law given by Osterbrock
(1989), $f(H\alpha)-f(H\beta)$=$-0.37$, $c$ can be readily determined
from Eq.(1).  Any corrected emission-line flux, $I_0(\lambda)$, can
then be estimated by correcting the extinction obtained from Eq.(1)
and the following average extinction law taken from Osterbrock (1989).
The extinction parameter $A_V$ ($V$ for visual) has been calculated
following the suggestion of Seaton (\cite{Seaton79}):
$A_V=E(B-V)R=\frac{cR}{1.47}$ (mag).  $R$ which is 3.2, is the ratio
of total to selective extinction at $V$.  The derived extinction
values of the sample galaxies, $A_V$, are given in Col.\,2 in
Table~\ref{tab3}.  Fig.~\ref{fig4} shows the distributions of galaxy
numbers vs. extinction ($A_V$) for the ``CFRS H$\alpha$-single",
``CFRS normal emission line" galaxies and the combination of these two
with a 0.5 bin in $A_V$.  ``CFRS H$\alpha$-single" galaxies mostly
display high extinction coefficients (median $A_V$=2.2), conversely to
``CFRS normal emission line" galaxies (median $A_V$=0.6).  However,
taking all the 16 galaxies together leads to an average extinction
coefficient of about $A_V$=1.25, which is equivalent to the assumption
of Fujita et al. (2003, $A_{H\alpha}$=1.0). 
No very high extinctions
($A_V>$ 3.3) are found in the sample galaxies.

\subsection{Photometry and dust correction}
Using the HST and the CFHT images of the sample galaxies, we used the
GIM2D{\footnote{GIM2D, Galaxy Image 2D, is an IRAF/SPP package written
to perform detailed bulge+disk surface brightness profile
decompositions of low signal-to-noise images of distant galaxies in a
fully automated way (Simard et al. \cite{simard02}).}}  software
package to calculate the inclination $i$ (the disk axis to the line of
sight) and the luminosities ratio of the bulge over the total (B/T) of
these galaxies.  Table~\ref{tab3} gives the corresponding values.  All
``CFRS H$\alpha$-single" galaxies show disk properties with B/T ratios
lower than 0.5.  The B/T values are consistent with the study of Kent
(\cite{Kent85}), which showed that the B/T ratio is mainly between 0.4
to 0.0 for Sab--Sc$^+$ galaxies (see their Fig.\,6) 
(also see Lilly et al. \cite{lilly98}).  Three ``CFRS
normal emission line" galaxies (CFRS03.0003, 14.1103 and 22.0474) are very
compact and the analysis of their CFHT images hardly recover their
morphological parameters (which results in large error bars).

 Generally, the internal extinction of galaxies increases with the
inclination of the disk because the path length through the disk
increases roughly as 1\,/cos$\,i$ (Giovanelli et al. \cite{Gio95}).
Fig.~\ref{fig6} provides a good illustration of this effect, since for
galaxies with inclination lower than 45{$^{\circ}$}, the median $A_V$
is 0.6 which could be compared to 2.2 in the edge-on galaxies.

   \begin{figure*}  
   \includegraphics[bb=80 318 510 810,width=18cm,clip]{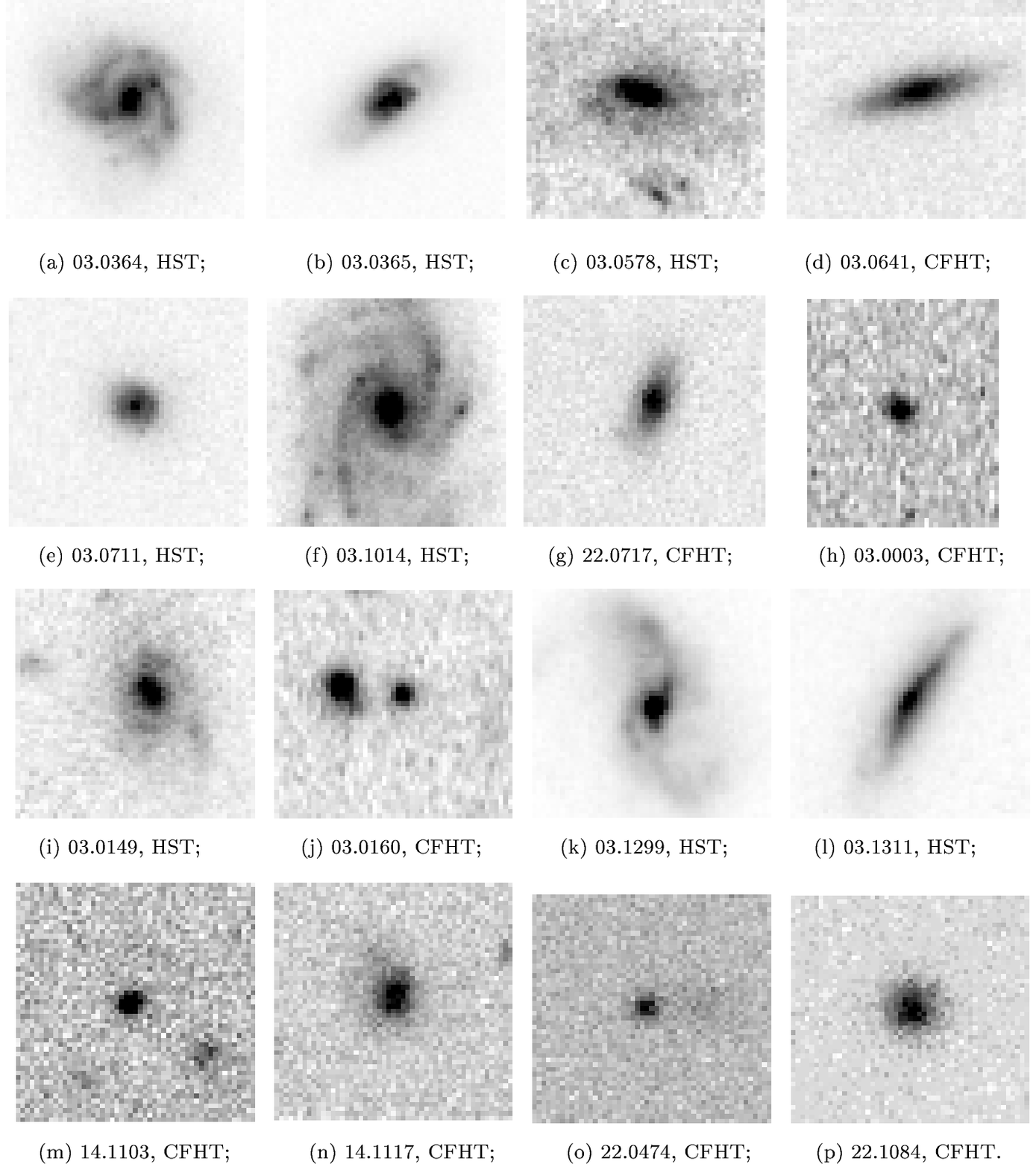}
 
   \caption {HST/F814W or CFHT images of the sample galaxies.  {\bf
 (a)-(g)} are the images of the seven ``CFRS H$\alpha$-single"
 galaxies; the rest nine are ``CFRS normal emission line" galaxies. Images
 are 5$^{\prime\prime}$$\times$5$^{\prime\prime}$ size, only 03.0003
 is 4$^{\prime\prime}$ in the horizontal size since it is near to the
 edge of the image of the field.}  
\label{fig5} 
\end{figure*}

{
\begin{table} 
{ \begin{center}  \tiny  
   \caption[] {\centering The photometric properties of the sample galaxies}  
 \label{tab3}
 \begin{tabular}{ccccc} \hline               
CFRS    &   $A_V$     &   B/T                & inclination $i$   \\ 
        &             &                      &  (degree) \\ \hline 
03.0364 & 2.84$\pm{0.21}$ & 0.025$_{-0.002}^{+0.002}$ & 11$_{-1}^{+8}$   \\ [1.5mm]
03.0365 & 2.13$\pm{0.26}$ & 0.010$_{-0.010}^{+0.024}$ & 59$_{-1}^{+1}$  \\ [1.5mm]
03.0578 & 2.33$\pm{0.44}$ & 0.214$_{-0.054}^{+0.047}$ & 55$_{-3}^{+4}$  \\ [1.5mm]
03.0641 & 1.22$\pm{0.36}$ & 0.007$_{-0.007}^{+0.037}$ & 84$_{-1}^{+1}$  \\  [1.5mm]
03.0711 & 0.14$\pm{0.14}$ & 0.003$_{-0.003}^{+0.013}$ & 32$_{-3}^{+4}$  \\ [1.5mm]
03.1014 & 2.41$\pm{0.41}$ & 0.021$_{-0.003}^{+0.157}$ & 53$_{-1}^{+4}$   \\ [1.5mm]
22.0717 & 2.75$\pm{0.39}$ & 0.095$_{-0.095}^{+0.089}$ & 72$_{-2}^{+2}$  \\ [1.0mm] \hline 
                          
03.0003 & 0.52$\pm{0.27}$ & 0.868$_{-0.434}^{+0.132}$ & 17$_{-17}^{+33}$  \\ [1.5mm]
03.0149 & 0.59$\pm{0.18}$ & 0.333$_{-0.074}^{+0.150}$ & 51$_{-3}^{+4}$    \\ [1.5mm]
03.0160 & 0.05$_{-0.05}^{+0.30}$ & 0.036$_{-0.036}^{+0.224}$ & 23$_{-23}^{+54}$    \\ [1.5mm]
03.1299 & 2.87$\pm{0.24}$ & 0.097$_{-0.007}^{+0.071}$ & 59$_{-1}^{+2}$  \\ [1.5mm]
03.1311 & 2.75$\pm{0.21}$ & 0.038$_{-0.013}^{+0.015}$ & 75$_{-1}^{+1}$ \\ [1.5mm]
14.1103 & 0.00$_{-0.00}^{+0.02}$ & 0.817$_{-0.817}^{+0.183}$ & 39$_{-13}^{+19}$ \\ [1.5mm]
14.1117 & 0.01$_{-0.01}^{+0.22}$ & 0.147$_{-0.147}^{+0.164}$ & 47$_{-13}^{+7}$  \\ [1.5mm]
22.0474 & 0.85$\pm{0.13}$ &  0.898$_{-0.030}^{+0.020}$ & 27$_{-3}^{+4}$  \\ [1.5mm]
22.1084 & 1.49$\pm{0.32 }$&  0.000$_{-0.000}^{+0.054}$  & 2$_{-2}^{+8}$    \\ \hline
\end{tabular} 
\end{center}
}
\end{table} 
}

\begin{figure}
  \includegraphics[bb=28 308 589 709,width=8.8cm,clip]{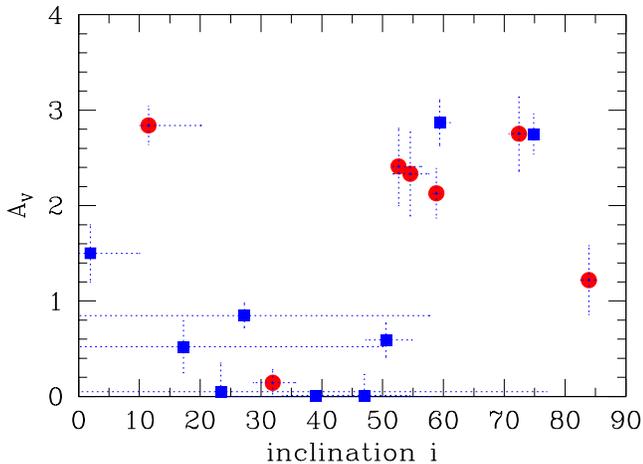}
 
\caption{The relations between extinction $A_V$ and inclination $i$ in
the sample galaxies.  Symbols as in Fig.~\ref{fig1}.}  
\label{fig6}
\end{figure}

\section{Diagnostic diagrams and gas abundances}
\subsection{Diagnostic diagrams}
\begin{figure} 
   \centering
  \includegraphics[bb=143 388 451 800,width=9.2cm,clip]{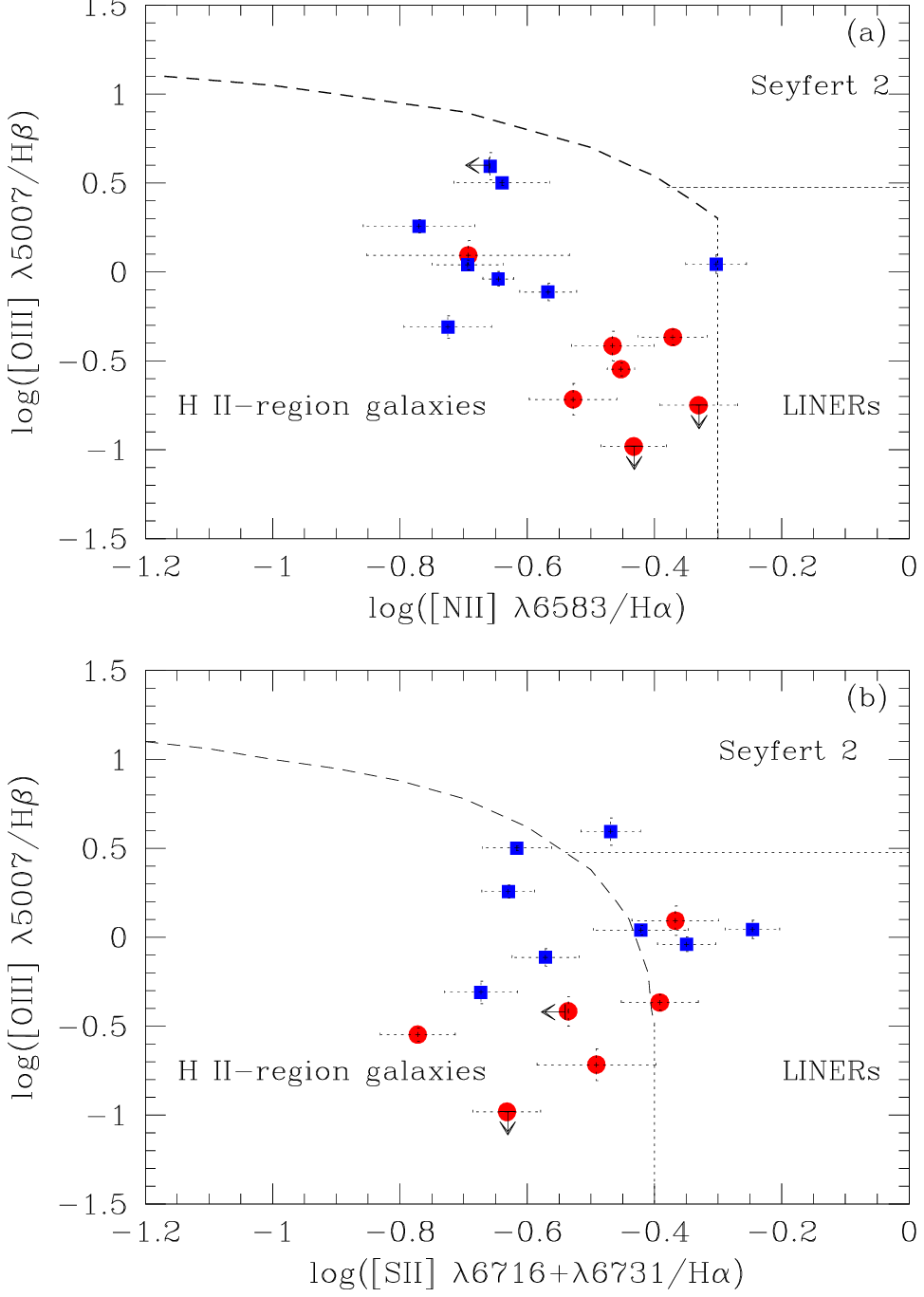}

  \caption{ {\bf (a)} \ion{[O}{iii]}/H$\beta$
vs. \ion{[N}{ii]}/H$\alpha$ diagnostic diagram for the sample
galaxies; {\bf (b)} \ion{[O}{iii]}/H$\beta$
vs. \ion{[S}{ii]}/H$\alpha$ diagnostic diagram for the sample
galaxies.  Symbols as in Fig.~\ref{fig1}.}  
\label{fig7} 
\end{figure}

Several emission line ratios have been used for a proper diagnostic
for the galaxies.  Fig.~\ref{fig7}(a) and (b) give the diagnostic
diagrams of log(\ion{[O}{iii]} $\lambda{5007}$/H$\alpha$) vs.
log(\ion{[N}{ii]} $\lambda{6583}$/H$\alpha$) and log(\ion{[O}{iii]}
$\lambda{5007}$/H$\alpha$) vs.  log(\ion{[S}{ii]}
$\lambda{6716}+\lambda{6731}$/H$\alpha$), respectively.

The \ion{[O}{iii]}/H$\beta$ ratio is mainly an indicator of the mean
level of ionization and temperature, while the \ion{[S}{ii]}/H$\alpha$
ratio is an indicator of the relative importance of a large partially
ionized zone produced by high-energy photoionization.  The
\ion{[N}{ii]}/H$\alpha$ ratio also gives a good separation between
H\,II region nuclei and Active Galactic Nuclei (AGN) though its
significance is not so immediately obvious.  The ratios have been
chosen to minimize the effects of dust extinction. (Veilleux \&
Osterbrock \cite{VO87}; Osterbrock \cite{Oster89})

The dashed and dotted lines in the two diagrams are the separating lines
between H\,II region nuclei and AGNs taken from Osterbrock
(\cite{Oster89}, their Fig.\,12.1 and 12.3).  The H\,II region-like
objects can be H\,II regions in external galaxies, starbursts, or
H\,II region galaxies, objects known to be photoionized by OB stars.
Seyfert 2 galaxies have relatively high ionization with \ion{[O}{iii]}
$\lambda{5007}$/H$\beta$$\geq$3. Most starburst and H\,II region
galaxies have lower ionization.  Many low-ionization galaxies have
stronger \ion{[S}{ii]} $\lambda$$\lambda$6716, 6731 and \ion{[N}{ii]}
$\lambda$6583 than H\,II regions or starburst galaxies. These objects
have been named ``Low-Ionization Nuclear Emission-line Regions"
(LINERs).

From Fig.~\ref{fig7}(a), it seems that all of the sample galaxies are
H\,II-region galaxies except for the possible ``LINER" property of
CFRS03.0160 due to the strong \ion{[N}{ii]} emission.  Fig.~\ref{fig7}(b)
confirms again that most of the studied galaxies lie in the
H\,II-region locus though few of them are in the active region locus
(LINER or Seyfert), including CFRS03.0160.  Shock-wave ionization may
produce stronger \ion{[S}{ii]} $\lambda\lambda{6716,6731}$ emissions 
relative to H$\alpha$ than in
typical H\,II regions.  For some of the best quality spectra, we have
been able to estimate the intensity ratio of the two \ion{[S}{ii]}
emission line,
(\ion{[S}{ii]}$\lambda{6716}$)/(\ion{[S}{ii]}$\lambda{6731}$) which
can be used to estimate the electron density, N$_e$ (Osterbrock
\cite{Oster89}, p134, their Fig.\,5.3). It results the values which
are generally close to what is expected from H\,II region galaxies
(may also see van Zee et al. \cite{Zee98}).

From the combination of Fig.~\ref{fig7}(a) and (b), we find that it
seems that only one object (CFRS03.0160) out of 16 could be a LINER.
Moderate resolution spectroscopy is required to establish these two
diagnostic diagrams, and is unique for solid estimates of the nature
of emission line objects. 
 As also noticed by Tresse et
al. (1996), a large fraction of their assumed Seyfert\,2 galaxies would
be better classified as LINERs if the underlying absorption under the
$H\beta$ line was properly accounted for.

In both Fig.~\ref{fig7}(a) and (b), the ``CFRS H$\alpha$-single"
galaxies and the ``CFRS normal emission line" galaxies lie in well
distinct areas, with the noticeable exception of CFRS03.0578.  We
believe this is related to different gas metal abundance histories in
these two classes of galaxies as it is studied in the following sections.

{
\begin{table*} 
{ \begin{center} 
 \tiny
\caption {The derived emission line ratios and oxygen, nitrogen
abundances of the sample galaxies.  The first seven are the ``CFRS
H$\alpha$-single" galaxies, and the rest nine are the ``CFRS normal
emission line" galaxies}
\label{tab4}

\begin{tabular}{rrrrrrrrrrr} \hline
 CFRS   & log$\ion{[O}{iii]}\over{{\rm H}\beta}$   &log$\ion{[N}{ii]}\over{{\rm H}\alpha}$ 
  & log$\ion{[S}{ii]}\over{{\rm H}\alpha}$   & log$\ion{[O}{ii]}\over{{\rm H}\beta}$  &
${\ion{[S}{ii]}_{1}}\over{\ion{[S}{ii]}_{2}}$
  & N$_e$  & log(R$_{23}$)  &    12+log(O/H)& t$_{\ion{[N}{ii]}}$ (K) &
log(N/O)  \\   
   & & & & & & (cm$^{-3}$)  &   & &     &
\\  \hline

03.0364    &  $-$0.55$\pm$0.04 & $-$0.45$\pm$0.02 
           & $-$0.77$\pm$0.06 &  0.30$\pm$0.30   & --- 
          &   ---      &   0.38$\pm$0.26 &    9.09$\pm$0.17 & 7082$\pm$1095 
           &$-$0.90$\pm$0.01 \\    [1.5mm]

03.0365  & $<$$-$0.98    &  $-$0.43$\pm$0.05 
         & $-$0.63$\pm$0.05 & 0.29$\pm$0.47 & 1.35$\pm$0.22 
         & $\sim$10$^2$ & $<$0.32   &  $>$9.13 & $<$6861  
         & $\sim$$-$0.86   \\ [1.5mm]

03.0578   &0.09$\pm$0.08 & $-$0.69$\pm$0.16
         &$-$0.37$\pm$0.07 & 0.68$\pm$0.22 & 0.87$\pm$0.27 
         & $\sim$9$\times$10$^2$   & 0.81$\pm$0.13 &  8.61$\pm$0.23 & 10074$\pm$1343 
        & $-$1.12$\pm$0.18 \\  [1.5mm]

03.0641  & $-$0.42$\pm$0.08 & $-$0.47$\pm$0.04   
         & $<$$-$0.54     & 0.17$\pm$0.24 &0.84$\pm$0.53 
         & $\sim$10$^3$  & 0.30$\pm$0.16 &  9.13$\pm$0.09 & 6798$\pm$571  
      & $-$0.82$\pm$0.19  \\ [1.5mm]

03.0711  & $-$0.37$\pm$0.04 & $-$0.37$\pm$ 0.05 
         &$-$0.39$\pm$0.06  & $-$0.24$\pm$0.37  & 1.81$\pm$0.54  
         & $<$10$^1$ & 0.06$\pm$0.20 &  9.24$\pm$0.07 & 6173$\pm$375  
        & $-$0.47$\pm$0.38  \\ [1.5mm]

03.1014 & $-$0.7$\pm$0.10 & $-$0.53$\pm$0.07 
         & $-$0.49$\pm$0.09   & 0.40$\pm$0.48 & 1.81$\pm$0.70 
         & $<$10$^1$  & 0.44$\pm$0.23 &  9.05$\pm$0.17 & 7384$\pm$1135 
        & $-$1.02$\pm$0.58  \\ [1.5mm]

22.0717  & $<$$-$0.75          & $-$0.33$\pm$0.06 
         & 0.07$\pm$0.05 & $-$0.08$\pm$0.49 & ---
         &  ---  & $<$0.03    &  $>$9.25   & $<$6111  
        & $\sim$$-$0.57 \\ [1.5mm]

 \hline   

03.0003  & 0.60$\pm$0.08 &  $<$$-$0.66 
         &  $-$0.47$\pm$0.05 &  0.01$\pm$0.34  & ---
         &  --- & 0.80$\pm$0.12  &  8.63$\pm$0.21 & $<$9965 
         & $<$$-$0.63 \\ [1.5mm]

03.0149  & 0.04$\pm$0.03 & $-$0.69$\pm$0.06  
         & $-$0.42$\pm$0.07 & 0.25$\pm$0.46  & 1.51$\pm$0.50 
         & $<$10$^1$     & 0.51$\pm$0.26 &  8.99$\pm$0.22 & 7756$\pm$1465 
       & $-$0.97$\pm$0.56 \\  [1.5mm]

03.0160   &   0.04$\pm$0.05 &  $-$0.30$\pm$0.05 
          & $-$0.25$\pm$0.04 & 0.49$\pm$0.31 &  ---
          & ---   &    0.66$\pm$0.22 &   8.83$\pm$0.28 & 8752$\pm$1720
          & $-$0.86$\pm$0.01\\ [1.5mm]

03.1299     & $-$0.11$\pm$0.05 & $-$0.57$\pm$0.04 
           &  $-$0.57$\pm$0.05 &  0.46 $\pm$0.31         &---
           & ---  & 0.60$\pm$0.24 &    8.91$\pm$0.25 &  8273$\pm$1621  
           & $-$1.02$\pm$0.01\\  [1.5mm]

03.1311   &  $-$0.04$\pm$0.04 & $-$0.64$\pm$0.02 
          & $-$0.35$\pm$0.05  & 0.50$\pm$0.30 & 1.58$\pm$0.20
           &  $<$10$^1$       &   0.64$\pm$0.23 &   8.86$\pm$0.27 &  8612$\pm$1707 
           &  $-$1.10$\pm$0.02\\  [1.5mm]

14.1103    &  0.81$\pm$0.00      & 9998
           &   9998    &  $<$$-$1.24     &---
           & ---        &  0.94$\pm$0.00 &  $<$7.63 & $<$11583 & ---\\  [1.5mm]

14.1117    &  0.26$\pm$0.04 & $-$0.77$\pm$0.09 
           & $-$0.63$\pm$0.04 & 0.54$\pm$0.31    &---
           & ---      &   0.77$\pm$0.19 &   8.68$\pm$0.30 &  9664$\pm$1782 
            & $-$1.17$\pm$0.03 \\ [1.5mm]

22.0474   &    0.50$\pm$0.02 & $-$0.61$\pm$0.07
          & $-$0.62$\pm$0.05 &  0.22$\pm$0.08  & 1.19$\pm$0.23 
           &$\sim$3$\times$10$^2$   & 0.77$\pm$0.03 &  8.68$\pm$0.05 &   9679 $\pm$285 
       & $-$0.72$\pm$0.06\\  [1.5mm]

22.1084   &  $-$0.31$\pm$0.06 & $-$0.72$\pm$0.07 
          & $-$0.67$\pm$0.06 &  0.41$\pm$0.32     & ---
           &  ---     &    0.51$\pm$0.26 &   8.99$\pm$0.23 &  7727$\pm$1507 
         & $-$1.18$\pm$0.01 \\  \hline
 \end{tabular} 

\end{center}
}

{\scriptsize Notes: 
log$\ion{[O}{iii]}\over{{\rm H}\beta}$=log$\ion{[O}{iii]}_{5007}\over{{\rm H}\beta}$,
 log$\ion{[N}{ii]}\over{{\rm H}\alpha}$=log$\ion{[N}{ii]}_{6583}\over{{\rm H}\alpha}$,
  log$\ion{[S}{ii]}\over{{\rm H}\alpha}$=log$\ion{[S}{ii]}_{6716+6731}\over{{\rm H}\alpha}$,
 \ion{[S}{ii]}$_{1}$=\ion{[S}{ii]}$_{6716}$, \ion{[S}{ii]}$_{2}$=\ion{[S}{ii]}$_{6731}$.
  }
\end{table*} 
}

\subsection{Metallicities from comparison with local H\,II regions and H\,II galaxies}
Gas abundance is a key factor to understand the star formation history
 and stellar population components of galaxies.  Metallicities of
 galaxies can be roughly estimated from the diagnostic diagram of
 \ion{[O}{ii]}/H$\beta$ vs. \ion{[O}{iii]}/H$\beta$ by comparing with
 the known metallicities of the local H\,II regions and H\,II galaxies
 (Hammer et al. \cite{hammer97}).

\begin{figure} 
   \includegraphics[bb=19 253 586 710,width=8.8cm,clip]{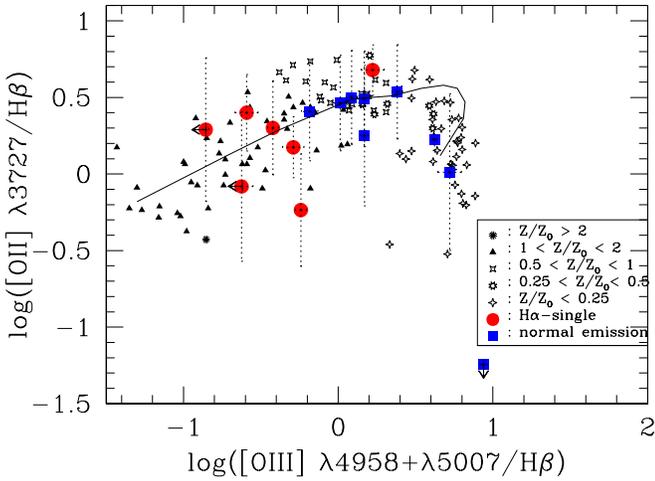}

   \caption{The relations between \ion{[O}{ii]}/H$\beta$ and
\ion{[O}{iii]}/H$\beta$ in the ``CFRS H$\alpha$-single" galaxies 
($filled ~ circles$) and the ``CFRS normal emission line" galaxies
($filled ~ squares$), together with a sample of the local H\,II
regions and H\,II galaxies with different metallicities. The solid
line shows the theoretical sequence from McCall et al. (\cite{MR85}),
which fits the local H\,II galaxies well, with metallicity decreasing from
the left to the right (also see Hammer et al. \cite{hammer97}).}
\label{fig8} 
\end{figure}

Figure~\ref{fig8} gives the log(\ion{[O}{ii]}
$\lambda{3727}$/H$\beta$) vs.  log(\ion{[O}{iii]}
$\lambda{4958}+\lambda{5007}$/H$\beta$) relations in ``CFRS
H$\alpha$-single" galaxies (the filled circles) and ``CFRS normal
emission line" galaxies (the filled squares), associated with a sample
of local H\,II regions and H\,II galaxies with known metallicities.
The data points of other galaxies are taken from the literature (see
Hammer et al. 1997 for references).  It shows that the ``CFRS
H$\alpha$-single" galaxies are more metal-rich than the ``normal
emission line" galaxies except CFRS03.0578, which results in weaker
\ion{[O}{ii]} and \ion{[O}{iii]} emissions.

Since \ion{[O}{ii]} emission lines are outside the rest-frame
wavelength ranges in six galaxies (``9999" were marked for their
fluxes of \ion{[O}{ii]} emission in Table~\ref{tab2}), we use the
theoretical \ion{[O}{ii]}/H$\beta$ values of the local H\,II regions 
and H\,II galaxies (the solid line on the figure) to be their
\ion{[O}{ii]}/H$\beta$ values to estimate the metallicities, which is
reliable by virtue of their H\,II-region galaxies properties certified
by Fig.~\ref{fig7} and the reliable \ion{[O}{iii]}/H$\beta$ ratios.
Actually, the \ion{[O}{iii]}/H$\beta$ ratio values have exhibited the
metallicities of these galaxies by comparing with the corresponding
ratios of the local H\,II regions and H\,II galaxies with different
metallicities on Fig.~\ref{fig8}.  Their error bars of \ion{[O}{ii]}
fluxes are estimated by using the average error of other sample
galaxies.


\subsection{Oxygen and nitrogen abundances}
Oxygen is one of the main coolants in the nebular occurring primarily
either via fine-structure lines in the far-infrared (52 and 88 $\mu$m)
when the electron temperature is low or via forbidden lines in the
optical (\ion {[O}{ii]} $\lambda{3727}$, \ion {[O}{iii]}
$\lambda{4958}$ and \ion {[O}{iii]} $\lambda{5007}$) when the electron
temperature is high.
 
Given the absence of reliable \ion{[O}{iii]} $\lambda{4363}$
detection, which is too weak to be measure except in extreme
metal-poor galaxies, to derive the electron temperature of the ionized
medium by comparing with \ion{[O}{iii]} $\lambda\lambda{4958,5007}$ lines
(Osterbrock 1989), oxygen abundances may also be determined from the
ratio of \ion{[O}{ii]}+\ion{[O}{iii]} to H$\beta$ lines (``strong
line" method). The general parameter is $R_{23}$: $ R_{23}=(\ion
{[O}{ii]} \lambda{3727}+\ion {[O}{iii]} \lambda{4958}+\ion {[O}{iii]}
\lambda{5007})/H\beta$.
To convert $R_{23}$ into 12+log(O/H), we adopt the analytical
approximation given by Zaritsky et al. ({\cite{ZKH94}}) (hereafter
ZKH), which is consistent with other calibration relations (see
Kobulnicky \& Zaritsky 1999, KZ99), and that is itself a polynomial
fit to the average of three earlier calibrations for metal rich H\,II
regions.
This relationship has been used for all the galaxies except for
CFRS14.1103 which presents a very small \ion{[O}{ii]}/$H\beta$ ratio and no
\ion{[N}{ii]} $\lambda\lambda{6548,6583}$ and \ion{[S}{ii]}
$\lambda\lambda{6716,6731}$ emission lines are detected which characterize 
a low oxygen abundance medium, and for which we adopt the relation from 
Kobulnicky et al. (1999) for metal-poor branch galaxies.  Its derived 
abundance by us is very similar to that of Tresse et al. (\cite{TH93}).

The derived oxygen abundances of the sample galaxies are given in
Table~\ref{tab4} as 12+log(O/H), which are consistent with the results
of Fig.~\ref{fig8}.  The ``CFRS H$\alpha$-single" galaxies have larger
abundance values than those of ``CFRS normal emission line" galaxies,
and lie in a region occupied by over-solar abundance objects, except
for CFRS03.0578.

\begin{figure}
  \includegraphics[bb=19 309 586 710,width=8.8cm,clip]{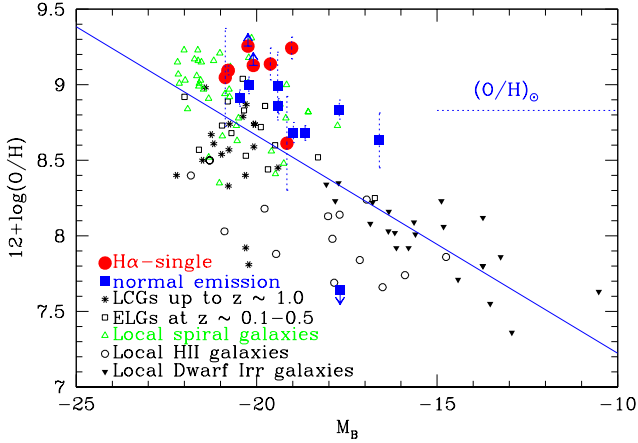}
\caption{Oxygen abundances vs. absolute blue magnitudes for the ``CFRS
H$\alpha$-single" galaxies ($filled ~ circles$) and the ``CFRS normal
emission line" galaxies ($filled ~ squares$), associated with the
Luminous Compact Galaxies (LCGs) up to $z$$\sim$1.0 ($asterisks$, from
Gruel \cite{Gruel02} and Hammer et al. \cite{hammer01}), the Emission
Line Galaxies (ELGs) at $z$$\sim$0.1-0.5 ($open$ $squares$, from
Kobulnicky \& Zaritsky \cite{KZ99}: KZ99), the Local Spiral Galaxies
($open$ $triangles$, from ZKH), the Local H\,II Galaxies ($open ~
circles$, from Telles \& Terlevich \cite{TT971}) and the Local Dwarf
Irregular Galaxies ($upside-down$ $filled$ $triangles$, from Richer \& McCall
\cite{RM95}).  All of them have been done using a spectral resolution
comparable to ours.  The solid line is a linear least-squares fit to
local irregular and spiral galaxies (from KZ99). Solar oxygen
abundance is shown by the dotted line.}
\label{fig9}
\end{figure}

Figure~\ref{fig9} shows the oxygen abundance vs. absolute blue
magnitude $M_B$ relations for the sample galaxies.  It shows that
the ``CFRS H$\alpha$-single" galaxies are lying in the top metallicity area of
local spiral galaxies and Emission Line Galaxies (ELGs) at $z=0.1\sim
0.5$ studied by KZ99, and the ``CFRS normal emission line" galaxies
are very similar to the local spiral galaxies except CFRS14.1103.  The
general trend of the sample galaxies follows: the brighter galaxies
are more metal-rich.  The Solar oxygen abundance
(12+log(O/H)$_{\odot}$=8.83) was taken from Grevesse \& Sauval
(\cite{GS98}).

\ion{[N}{ii]} $\lambda$6583 can be used in conjunction with
\ion{[O}{ii]} $\lambda$3727 and the temperature in the \ion{[N}{ii]}
emission regions (t$_{\ion{[N}{ii]}}$) to estimate the N/O ratio in
the sample galaxies assuming $\frac{\rm N}{\rm O}=\frac{\rm N^{+}}{\rm
O^{+}}$.  Uncertainties due to emission line measurements, reddening
and sky subtraction in the presence of strong night sky emission lines
near \ion{[N}{ii]} dominate the error budget for N/O.

\begin{figure}
   \includegraphics[bb=19 309 586 710,width=8.8cm,clip]{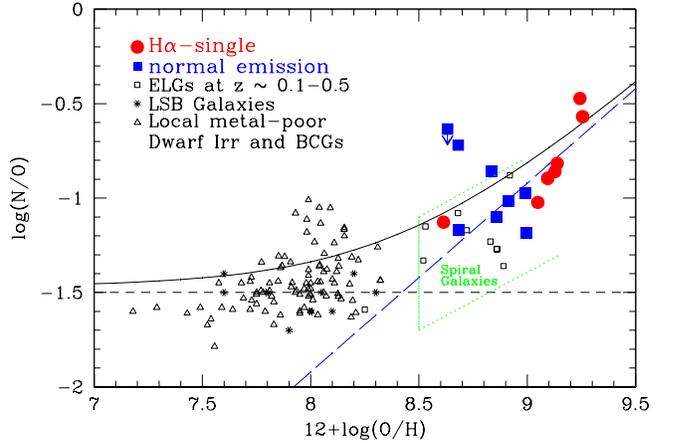}

\caption{ N/O vs. O/H relations for the ``CFRS H$\alpha$-single"
galaxies ($filled ~ circles$) and the ``CFRS normal emission line"
galaxies ($filled ~ squares$), associated with the ELGs at
$z$$\sim$0.1-0.5 ($open$ $squares$, from KZ99), the Low Surface
Brightness (LSB) Galaxies ($asterisks$, from van Zee et al.
\cite{Zee97}), the Local metal-poor Dwarf Irregular and Blue Compact
Galaxies (BCGs) ($open$ $triangles$, from Kobulnicky \& Skillman
\cite{KS96} and Izotov \& Thuan \cite{IT99}).  All of them have been
done using a spectral resolution comparable to ours.  The dotted box
for spiral galaxies is taken from KZ99.  The horizontal dot-dashed
line refers to the $primary$ origin of nitrogen, the sloping dashed
line refers to the $secondary$ origin, and the solid line is the
combined contribution of a (delayed) $primary$ and a $secondary$
component of nitrogen (taken from Vila-Costas \& Edmunds
\cite{VilaE93}).  }
\label{fig10}
\end{figure}

First, we use the formula given by Thurston et al. (\cite{TEH96}) to
estimate the temperature in the \ion{[N}{ii]} emission region
(t$_{\ion{[N}{ii]}}$) by using log$R_{23}$.  Then, log(N/O) is
estimated from (\ion{[N}{ii]}
$\lambda{6548}+\lambda{6583}$)/(\ion{[O}{ii]} $\lambda {3727}$)
emission ratio and t$_{\ion{[N}{ii]}}$.  The derived values are given
in Table~\ref{tab4}.  Fig.~\ref{fig10} gives the log(N/O)
vs. 12+log(O/H) relations for these sample galaxies.  Most of them
follow the $secondary$ nitrogen production well (Vila-Costas \&
Edmunds \cite{VilaE93}).

\section{ H$\alpha$ Luminosities and star formation rates}
\subsection{H$\alpha$ luminosities and SFRs}
There is unique advantage at using H$\alpha$ to obtain the SFRs for
low-$z$ galaxies.  Among the Balmer lines, H$\alpha$ is the most
directly proportional to the ionizing UV flux, and the weaker Balmer
lines are much more affected by stellar absorption and reddening.  The
SFRs of the ``CFRS H$\alpha$-single" galaxies were estimated from the
H$\alpha$ luminosities.

In the following, we adopt Salpeter initial mass function (IMF) 
with low and high mass cutoffs
at 0.1 and 100 $M_{\odot}$ (Salpeter \cite{Sal55}).  The calibrations of Kennicutt et
al. (\cite{K94}) and Madau et al.  (\cite{Madau98}) yield:

\begin{equation}
SFR (M_{\odot} yr^{-1})=7.9 \times 10^{-42}L(H\alpha)(ergs~ s^{-1}) \times Aper,
\end{equation}
(Kennicutt \cite{K98}), with
\begin{equation}
L(H\alpha)=4\pi(3.086 \times 10^{24}D_L)^2f(H\alpha)(ergs~ s^{-1}),
\end{equation} 
where $L(H\alpha)$ is the H$\alpha$ luminosity in ergs\,~s$^{-1}$,
$f(H\alpha)$ is the integrated flux in ergs\,~s$^{-1}$\,cm$^{-2}$
after correcting for the extinction, and $D_L$ is the luminosity
distance in Mpc.  $Aper$ is the aperture correction factor by
comparing the photometric and spectral magnitudes in $I_{AB}$ band due
to the limited size of the slit.  The related results of these
parameters and the derived SFRs are given in the left part of
Table~\ref{tab5} (Col.(1)-(5)), in which Flux$_c$(H$\alpha$) is the
H$\alpha$ emission flux after correcting for the extinction.  CFRS
galaxies have SFRs ranging from Milky Way value to higher typical
values of starburst galaxies.

{
\begin{table*} 
{\scriptsize
\centering 

\caption {Some derived characteristic parameters of the ``CFRS
H$\alpha$-single" (the top seven) and the 
``CFRS normal emission line" (the bottom nine) galaxies, the
left part displays the results derived from the higher resolution
spectra from the VLT or the CFHT, and the right part displays the
results from low resolution CFRS spectra.  The sequent columns in the
right part of the table show the measured fluxes of (H$\alpha$+\ion
{[N}{ii]}) emission line, the rest-frame EW values of (H$\alpha$+\ion
{[N}{ii]}) emission line (in units of \AA), the $N_2$ parameter, the
measured fluxes of H$\beta$, the aperture correction factor, the SFRs
with the average extinction $A_V=1$ (SFR$_1$), the estimated $A_V$
from the CFRS spectra (quoted as $A_{VC}$) and the derived SFRs by
using $A_{VC}$ (SFR$_C$).  All fluxes of the lines are given in units 
of (10$^{-17}$ ergs cm$^{-2}$ s$^{-1}$).  All SFRs are in units of
($M_{\odot}$ $yr^{-1}$)}
 
 \label{tab5}

 \begin{tabular}{ccccc|ccccccccccc} \hline
  & Moderate &  & resolution &  &       &  & Very& low & resolution &  & &  \\ [1.3mm]
        &  VLT &  or   & CFHT  &       & & & &  CFRS & & & &  \\  [1.3mm]   \hline      
CFRS     & Flux$_c$  & Aper  & L(H$\alpha$) &  SFR$~~~~$   &   Flux & REW & $N_2$ & Flux &
Aper:    &SFR$_1$  & $A_{VC}$ & SFR$_{\rm C}$     \\
          & (H$\alpha$) &     & ergs s$^{-1}$ & M$_{\odot} yr^{-1}$ &  (H$\alpha$+ \ion {[N}{ii]}) &  & & (H$\beta$) &
10$^{0.4a}$ &  $A_V$=1   &   & \\ 

 (1) & (2) & (3) & (4) & (5) & (6)      & (7) & (8)& (9) & (10) & (11) & (12)& (13) \\ [1.3mm]\hline

03.0364  &  608.9$\pm{100.8}$  &   2.14 & 10$^{42.26}$ &   31.22$\pm{5.17}$   &    94.8  & 44.5    &  0.42   &  0.0  &  2.14  &   7.52    & ---  &---                      \\  [1.8mm] 
03.0365 & 491.3$\pm{102.3}$ & 1.92  & 10$^{42.04}$  &16.84$\pm{3.50}$          &   121.8   & 52.6    &  0.36   & 0.0 &   1.49  &   5.22    &  ---  &---               \\[1.8mm]
03.0578 & 50.1$\pm{20.2}$  & 3.39  & 10$^{41.11}$  & 3.48$\pm{1.23}$          &    18.9   & 46.8    &  0.42   & 0.0 &   2.03  &   1.05    & ---    &---             \\[1.8mm]
03.0641 & 27.7$\pm{8.0}$  & 1.51  & 10$^{40.96}$  & 1.08$\pm{0.31}$          &    25.5   & 20.5    &  0.45   & 0.0 &   1.50  &   1.50    &  ---  & ---             \\ [1.8mm]
03.0711 & 15.2$\pm{1.82}$  & 1.37  & 10$^{40.70}$  & 0.54$\pm{0.06}$          &    14.4   & 32.9    &  0.45   & 0.0 &   1.28  &   0.73    &  ---  &---              \\[1.8mm]
03.1014 & 299.6$\pm{97.7}$  & 4.58  & 10$^{41.73}$  & 19.60$\pm{6.40}$         &   103.8   & 24.7    &  0.45   & 0.0 &   1.89  &   4.24    &  ---  & ---               \\ [1.8mm] 
22.0717 & 271.2$\pm{85.2}$  & 1.15  & 10$^{42.01}$  & 9.31$\pm{2.93}$          &    19.5   & 10.8    &  0.54   & --- &   1.41  &   1.17    & ---   &---              \\[1.8mm] \hline

03.0003 &  21.0$\pm{4.5}$ & 3.38  & 10$^{40.68}$  &  1.27$\pm{0.27 }$       &     8.2   & 95.0   &  0.21   &  1.7 &  1.22  &   0.34    & 0.90  & 0.21               \\[1.8mm]
03.0160  &   12.9$\pm{3.3}$   &  0.77  & 10$^{40.47}$ &    0.18$\pm{0.05}$   &    16.9   &38.9    &  0.45   &  1.2  &  0.77  &   0.35    & 3.13 &1.99                     \\  [1.8mm]  
03.0149 & 27.7$\pm{4.1}$ & 2.87  & 10$^{40.92}$   & 1.90$\pm{0.28}$          &    17.8   & 51.0    &  0.36   & 3.2 &   2.56  &   1.76    &  0.91  &  1.67          \\  [1.8mm]
03.1299  & 1442.1$\pm{281.0}$  &   2.28 & 10$^{42.31}$ &   37.18$\pm{7.24}$   &   224.0  & 77.0    &  0.36   & 16.1  &  2.28  &   9.33    & 3.25 & 58.2                    \\  [1.8mm] 
03.1311  &  531.2$\pm{88.0 }$   & 1.77  & 10$^{41.88}$ &   10.68$\pm{1.77}$   &    69.5  & 41.6    &  0.45   &  2.7  &  1.77  &   2.12    & 4.66 & 41.0                    \\  [1.8mm] 
14.1103   &  133.9$\pm{3.0}$   &  2.08  & 10$^{41.44}$ &    4.50$\pm{0.10}$   &    97.7  &2434.0   &  0.15   &  39.0 &  2.08  &   ---     & 0.00 &  2.86                   \\  [1.8mm] 
14.1117   &  16.0$\pm{2.8}$   &  4.73  & 10$^{40.44}$ &    1.03$\pm{0.18}$   &    19.6  & 69.0    &  0.36   &  4.7  &  4.73  &   2.04    & 0.18 &  1.07                   \\  [1.8mm] 
22.0474 & 102.7$\pm{10.9}$  & 1.10  & 10$^{41.59}$  &  3.40$\pm{0.36 }$        &    53.5   &312.0    &  0.15   & 10.4 &  1.54  &   4.77    & 1.14  & 5.40              \\[1.8mm]
22.1084   &  89.8$\pm{20.8}$  &   1.73  & 10$^{41.57}$ &    5.15$\pm{1.32}$   &    31.5  & 37.7    &  0.45   &  2.3  &  1.73  &   2.74    & 3.05 &  14.5                   \\    \hline

 \end{tabular} 
}
\end{table*} 
}

\subsection{Comparing the SFRs with those from low-resolution CFRS spectra}
To understand more the effect of spectral resolution on the derived
SFRs, we also estimated the SFRs of these galaxies from their
low-resolution CFRS spectra.  For the latter estimates, we have
followed the method suggested by TM98 and used the extinction law from
Osterbrock (1989), then the dereddened and aperture corrected
H$\alpha$ fluxes are estimated by:

\begin{equation}
f(H{\alpha})=\frac{f(H{\alpha}+ \ion {[N}{ii]})}{1+N_2}10^{c(1.13-0.37)}10^{0.4a},
\end{equation}
where $c=A_V\times1.47/3.2$ (see Sect.4). For the seven``CFRS
H$\alpha$-single" galaxies, extinction cannot be estimated from Balmer
decrement due to the absence of H$\beta$ emission, we use the average
$A_V=1$, corresponding to $c=0.45$, by following TM98.  $N_2$ is the
parameter to reflect \ion {[N}{ii]} $\lambda$6583 emission mixed with
H$\alpha$ in the CFRS spectra.  We got $N_2$ values from Fig.\,3(b) of
TM98 by considering the rest-frame EW(H$\alpha$+\ion {[N}{ii]}). $a$
refers to the aperture correction factor by comparing the photometric
and spectral magnitudes in $V_{AB}$ band.  Corresponding values are
given in the right part of Table~\ref{tab5} (Col.(6)-(13)).
 
\begin{figure}
\input epsf
\hspace {0cm} \epsfxsize 8.2cm \epsfbox{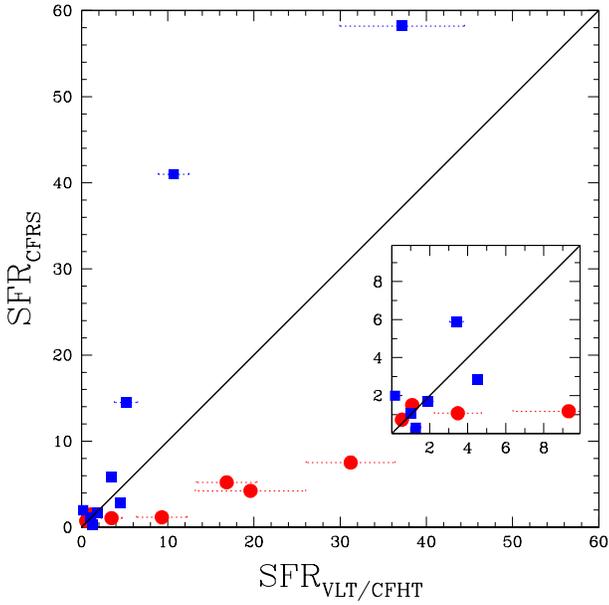}

\caption {Comparison between the SFRs of the sample galaxies obtained from
the low-resolution CFRS spectra (SFR$_{\rm CFRS}$) and the higher
quality spectra (SFR$_{\rm VLT/CFHT}$) studied in this work.  The
small figure on this figure shows the detail around the origin.  For
SFR$_{\rm CFRS}$, SFR$_1$ were used for the seven ``CFRS
H$\alpha$-single" galaxies, and SFR$_{\rm C}$ were used for the nine
``CFRS normal emission line" galaxies (see Table~\ref{tab5}).  Symbols
as in Fig.~\ref{fig1}.}

 \label{fig11}
     \end{figure}

Figure~\ref{fig11} shows how misleading are the SFR estimates based on
low resolution spectra by comparing the corresponding SFR$_{\rm CFRS}$
with the SFR$_{\rm VLT/CFHT}$ from the moderate resolution spectra.
Thus, it may be that TM98 systematically underestimated the SFRs of
the ``CFRS H$\alpha$-single" galaxies simply because they
underestimated the actual extinction coefficients of these galaxies.
Conversely, the SFRs of ``CFRS normal emission line" galaxies are
often overestimated since the underlying absorption beneath $H\beta$
could not be properly accounted for in low resolution spectroscopy and
leads to a severe overestimation of the extinction coefficient
properties. 
This effect probably generates the derived $A_{V}$ values exceeding 3
or 4 (see Table~\ref{tab5}, also Tresse et al. 1996).  
SFRs of individual galaxies can be only
recovered by a proper analysis of the higher quality spectra
(SFR$_{\rm VLT/CFHT}$) at moderate spectral resolution.

\subsection{Estimates of Cosmic Star Formation Density (CSFD)}

From the above, some qualitative arguments can be used to test the
validity of previous works based on low resolution spectroscopy or
narrow band filter imagery.  Although the latter cannot provide
quantitative SFR measurements of individual galaxies, it is valuable
to notice that, the SFR overestimates and underestimates are almost
balanced in the TM98 study (Fig.~\ref{fig11}).  Table~\ref{tab6}
report the total SFR budget assuming that the seven ``CFRS
H$\alpha$-single" galaxies and the nine ``CFRS normal emission line"
galaxies are the representative of the whole CFRS sample at low
redshift.  The difference between the two estimates is only 13\% which
is far below the error bars in TM98. 
Table~\ref{tab5} shows the extinction coefficients of 
``CFRS normal emission line" galaxies could be much overestimated 
from the very low resolution spectroscopy.
If unrealistic values of $A_{V}$ ($\geq$3.5) are taken to 
calculate the SFRs of these galaxies, hence the SF density,
this could lead to severe
overestimations. Recall that for the luminous infrared
galaxies, Flores et al. (2003) never find $A_{V}$ values larger than
3.5.  And the average color excesses, $E(B-V)$, of luminous infrared
galaxies (LIGs) studied by Veilleux et al. (\cite{VK95}) are only
0.72, 0.99 and 1.14 in Seyfert 2, H\,II LIGs and LINERs, respectively.
  
Fujita et al. (2003) have corrected their $H\alpha$ luminosities 
from narrow band filter imagery using
$A_{H\alpha}$= 1 which grossly corresponds to $A_{V}$ $\sim$
1.25. This value is in agreement with our median value for the 16
galaxies studied here.  However this correction is related to the
power law of the extinction coefficient, leading to important effect
related to the large extinction coefficients.  Table~\ref{tab7}
compares the effect of applying the Fujita et al. (2003) correction on
the 16 galaxies studied here.  The result is that Fujita et al. (2003)
might have underestimated their SF density by a factor close to 2.

It is out of the scope of this paper to provide a quantitative
estimates of the CSFD, because of the small number of objects
studied. Indeed, the study here would not help in reconciling the
different estimates at low redshift. It however strongly calls for a
systematic survey at moderate resolution of a complete sample of
galaxies detected from deep narrow band imagery, in order to correct
the $H\alpha$ luminosities by properly estimating the extinction
coefficients. 

{
\begin{table*}
{\begin{center}
   \caption[] {The comparison of the total SFRs
(in units of $M_{\odot}$ $yr^{-1}$) for the CFRS low-$z$ sample.
 The adopted interstellar extinction law
assumes ($f(H\alpha)-f(H\beta)$=$-$0.323) and the
total galaxies are 110 here by following TM98 }
  \label{tab6}
 \begin{tabular}{ccc} \hline
                              &   Total SFR$_{\rm VLT/CFHT}$  &  Total SFR$_{\rm CFRS}$  \\  \hline

57 ``CFRS H$\alpha$-single" galaxies & 696.5 &  160.3  \\
53  ``CFRS normal" galaxies     &      389.7  & 780.2  \\
Total 110                       & 1086.2     &   940.5 \\ \hline

 \end{tabular}
\end{center}}
\end{table*}
}

{
\begin{table*} 
{\begin{center} 
   \caption[] {The comparison of the SFRs (in units of $M_{\odot}$ $yr^{-1}$) 
of the 16 CFRS low-$z$ galaxies from different extinctions }  
 \label{tab7}
 \begin{tabular}{ccc} \hline               
       &   Total SFRs$_{\rm VLT/CFHT}$  &  Total SFRs$_{\rm VLT/CFHT}$  \\ 
       &    with well determined $A_V$ &   assuming $A_{H\alpha}=1$  \\  \hline

seven ``CFRS H$\alpha$-single" galaxies &  82.1            & 31.4   \\
nine  ``CFRS normal" galaxies           &    65.3          &  43.4 \\ 
Total 16                                &   147.4          &  74.8         \\ \hline
\end{tabular} 
\end{center}}
\end{table*} 
}

\begin{figure} 
   \includegraphics[bb=18 150 588 706,width=8.8cm,clip]{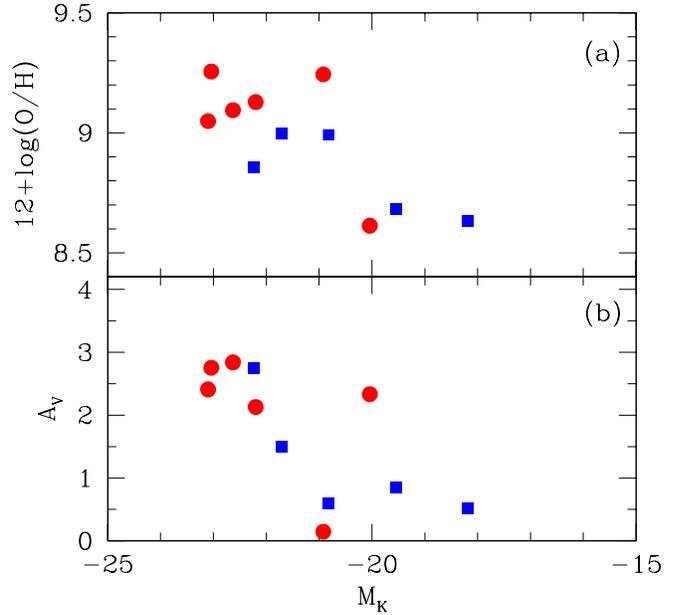}

   \caption{ Oxygen abundances (top) and extinction coefficients
(bottom) against absolute K band magnitudes for the sample
galaxies. Both relations show the trends that the higher abundances and
extinctions are in the more massive stellar systems. Symbols as in
Fig.~\ref{fig1}. } 
\label{fig12} 
\end{figure}
 

%
\section{Discussion and conclusion}

Using moderately high resolution (R$>$600) and high S/N spectra
obtained from the VLT and the CFHT, we have studied the properties of
a sample of 16 CFRS low redshift galaxies. This sample could be
splitted in seven ``CFRS H$\alpha$-single" emission galaxies, and nine
``CFRS normal emission line" galaxies, from their spectral properties
at the CFRS very low spectral resolution.  Selected from the CFRS
sample, these can be taken as representative of the H$\alpha$-emission
field galaxy population at $z$$\le$0.3.

Using the Balmer decrement method (H$\alpha$ to H$\beta$), we have
been able to calculate their interstellar extinction values, by
properly accounting for the underlying stellar absorption.

Two diagnostic diagrams (log(\ion {[O}{iii]} $\lambda{5007}$/H$\beta$)
vs. log(\ion {[N}{ii]} $\lambda{6583}$/H$\alpha$) and log(\ion
{[O}{iii]} $\lambda{5007}$/H$\beta$) vs.  log(\ion {[S}{ii]}
$\lambda6716+\lambda{6731}$/H$\alpha$) have been obtained to derive
firm conclusion about the nature of the emission line activity,
especially because that \ion {[N}{ii]} emission lines are divided from
H$\alpha$ emission in these higher resolution spectra.  Derivation of
extinction properties have allowed to accurately estimate their oxygen
and nitrogen abundances, as well as to calculate their SFRs using the
extinction corrected H$\alpha$ luminosities.

We find that the spectral properties of galaxies at very low spectral
resolution can be well understood by the properties of their
interstellar media.  Namely ``CFRS H$\alpha$-single emission" galaxies
shows systematically larger extinction coefficient, higher oxygen and
nitrogen abundances than the rest of the sample.  These properties
suffice to explain why $H\beta$ and \ion {[O}{iii]} $\lambda$5007
emissions are undetected by low resolution spectroscopy.  They can be
considered to be the mature and massive spirals that lie at low
redshifts as it can be derived from their K absolute magnitudes
(Fig.\ref{fig12}).

SFRs of these low redshift galaxies are ranging from 0.5 $M_{\odot}$
$yr^{-1}$ (Milky Way value) to 40 $M_{\odot}$ $yr^{-1}$ (strong
starburst). We also find that the SFRs of individual
galaxies cannot be properly derived using low resolution spectroscopy.
Indeed, extinction corrections are often large and requires a proper
account of the underlying stellar absorption to the Balmer lines, 
which is
simply impossible at spectral resolution lower than 600.  Previous
studies of SF density at low redshifts have assumed average
properties for underlying absorption or even for extinction of Balmer
line fluxes derived from low resolution spectroscopy. Indeed, $H\beta$
line is affected by underlying absorption and extinction in such a
complex way that only moderate resolution can estimate properly the
$H\alpha/H\beta$ ratio. Hence, the previous studies may systematically
underestimated the contribution of ``CFRS H$\alpha$-single" emission
galaxies (the mature and massive systems) and overestimated the
contribution of other normal emission line galaxies.

From the data shown here, one can only speculate about the consequences
at higher redshifts. Deep surveys preferentially select luminous
galaxies in their highest redshift bins, which generally show
relatively large extinction coefficients (see Fig.~\ref{fig12}).  
This effect may explain most of the $\sim$60\% gap between the SF
density derived by Tresse et al. (2002) 
(from $H\alpha$ luminosity, assuming $A_{V}$= 1) 
and that of Flores et al. (1999) (from combination of IR and UV measurements). 
Indeed, in a forthcoming paper, Flores et al. (2004, in preparation) find
that one third of the Tresse et al.'s sample are 
luminous infrared galaxies,
for which $A_V$ should reach values much larger than 1.

The present study gives a obvious warning for the studies based on low
resolution spectroscopy aimed at measuring individual galaxy
properties (gas chemical abundances, interstellar extinction, stellar
population, ages as well as star formation rates and history),
particularly for the metal rich and dusty spiral galaxies.  Because
this affects a large fraction of the galaxies, deriving cosmological
star formation density from low resolution spectroscopic surveys could
lead to severe biases.

\section*{Acknowledgments} 
We thank the referee for the valuable suggestions which leads us to
improve a lot this study.  We thank Dr. Rafael Guzm\'{a}n, Dominique
Proust and Jing-Yao Hu for the useful discussions and help. We also
thank Dr. Claude Carignan and Mark Neeser for the valuable suggestions
and the help to improve the English description.

\end{document}